\documentclass[preprint,12pt]{elsarticle}

\usepackage{graphicx}%
\usepackage{multirow}%
\usepackage{arydshln}
\usepackage{hhline}
\usepackage{amsmath,amssymb,amsfonts}%
\usepackage{amsthm}%
\usepackage{mathrsfs}%
\usepackage[title]{appendix}%
\usepackage{xcolor}%
\usepackage{textcomp}%
\usepackage{manyfoot}%
\usepackage{booktabs}%
\usepackage{algorithm}%
\usepackage{algorithmicx}%
\usepackage{algpseudocode}%
\usepackage{listings}%
\usepackage{url}

\usepackage{amssymb}

\journal{Computerized Medical Imaging and Graphics}

\begin{document}

\begin{frontmatter}



\title{Weakly supervised segmentation of intracranial aneurysms using a novel 3D focal modulation UNet}


\author[inst1]{Amirhossein Rasoulian} \ead{ah.rasoulian@gmail.com}
\author[inst1]{Arash Harirpoush}
\author[inst1]{Soorena Salari}
\author[inst1]{Yiming Xiao} \ead{yiming.xiao@concordia.ca}

\affiliation[inst1]{
            organization={Department of Computer Science and Software Engineering, Concordia University},
            city={Montreal},
            state={Quebec},
            country={Canada}
            }



\begin{abstract}
\noindent
Accurate identification and quantification of unruptured intracranial aneurysms (UIAs) is crucial for the risk assessment and treatment of this cerebrovascular disorder. Current 2D manual assessment on 3D magnetic resonance angiography (MRA) is suboptimal and time-consuming. In addition, one major issue in medical image segmentation is the need for large well-annotated data, which can be expensive to obtain. Techniques that mitigate this requirement, such as weakly supervised learning with coarse labels are highly desirable. In the paper, we propose FocalSegNet, a novel 3D focal modulation UNet, to detect an aneurysm and offer an initial, coarse segmentation of it from time-of-flight MRA image patches, which is further refined with a dense conditional random field (CRF) post-processing layer to produce a final segmentation map. We trained and evaluated our model on a public dataset, and in terms of UIA detection, our model showed a low false-positive rate of 0.21 and a high sensitivity of 0.80. For voxel-wise aneurysm segmentation, we achieved a Dice score of 0.68 and a 95\% Hausdorff distance of $\sim$0.95 mm, demonstrating its strong performance. We evaluated our algorithms against the state-of-the-art 3D Residual-UNet and Swin-UNETR, and illustrated the superior performance of our proposed FocalSegNet, highlighting the advantages of employing focal modulation for this task. The code will be available upon acceptance at \url{https://github.com/ah-rasoulian/FocalSegNet}.

\end{abstract}




\end{frontmatter}


\section{Introduction}
    \label{sec:intro}
    An intracranial aneurysm is an abnormal focal bulging or ballooning of the vasculature in the brain due to weakened blood vessel walls. Although often asymptomatic, large aneurysms can cause headaches, visual disturbance, neurological deficits, and other issues \cite{Frösen2012}. Furthermore, a ruptured aneurysm can lead to life-threatening subarachnoid hemorrhage, a severe stroke subtype. Typical treatments for unruptured intracranial aneurysms (UIAs) include endovascular coiling and surgical clipping. When screening for UIAs and determining the treatment plans, risk assessment of UIA growth and rupture is crucial. Thanks to modern medical imaging techniques, such as computed tomography (CT) and magnetic resonance imaging (MRI), the ability to accurately and reliably identify and measure aneurysms can ensure therapeutic outcomes. Compared with  CT angiography (CTA), time-of-flight magnetic resonance angiography (TOF-MRA) does not expose the patients to radiation or adverse reactions toward contrast agents. It thus is better suited for routine follow-up imaging.
    
    Manual identification and measurements of UIAs can be difficult and time-consuming, especially for those that are of small sizes ($<$5mm), and it is estimated that ~10$\%$ of all UIAs are missed during screening \cite{White2000}. Therefore, automatic algorithms that allow detection and especially 3D segmentation of the aneurysms from medical scans can greatly facilitate the clinical workflow and enable more fine-grained risk analysis based on UIA shapes. To date, many techniques \cite{Timmins2021,Din2023} have been proposed for UIA detection, but only a few reported performance for 3D UIA segmentation from CTA/MRA. Notably, the ADAM Challenge \cite{Timmins2021} hosted in conjunction with MICCAI 2020 attracted nine DL-based UIA segmentation algorithms. Among these, various approaches of convolutional neural networks (CNNs), especially variants of the UNet have been proposed, with the winning algorithm achieving an average Dice score of 0.64 and a mean 95$\%$ Hausdorff distance (95-HD) of 2.62mm for correctly detected UIAs. However, most existing techniques were developed based on private clinical data with refined manual annotations, and many focused on CTA. Further investigation is still needed for this challenging task, especially for MRA with weaker contrast. The lack of large well-annotated databases, which are costly to acquire and demand high domain expertise, poses significant challenges in developing deep learning methods in medical imaging applications, particularly for anatomical segmentation. To mitigate this, weakly supervised segmentation \cite{Rajchl2017,Yang2020,rasoulian2022weakly,melba:2023:012:rasoulian} leverages coarse annotations that are easier to obtain (e.g., rough segmentation and categorical image labels) to derive refined segmentation. In this paper, we employ coarse segmentation ground truths of UIAs from TOF-MRA to detect and segment the pathology by using a novel 3D focal modulation UNet, called FocalSegNet, and post-processing with a fully connected conditional random field (CRF) model. In summary, our work has three major novel contributions. \textbf{First}, inspired by the recent focal modulation technique \cite{Yang2022FocalMN} and the UNet architecture \cite{iek20163DUL}, we proposed a novel 3D UNet with focal modulation; \textbf{Second}, we thoroughly compared the performance of self-attention \cite{Vaswani2017AttentionIA} and focal modulation \cite{Yang2022FocalMN} in weak segmentation of UIAs; \textbf{Lastly}, we revealed the key factors that contribute to UIA segmentation for the proposed method through various ablation studies.

\section{Methodology}
    \label{sec:methodology}
    \subsection{Network architecture of FocalSegNet}
        \label{sec:methodology:arch}
\begin{figure}[t]
        \includegraphics[width=\textwidth]{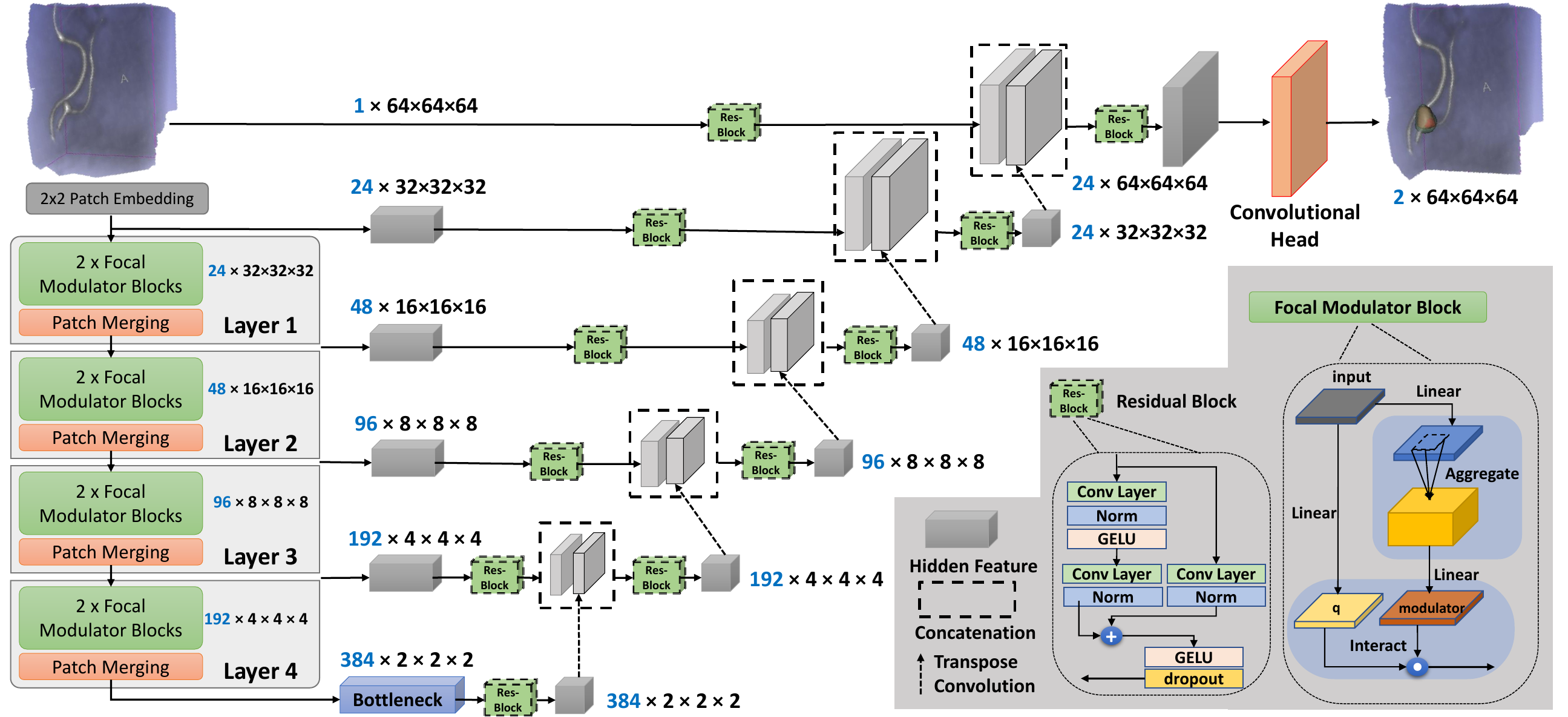}
        \caption{Network architecture of the proposed FocalSegNet} \label{focalconvunet}
        \end{figure}

        While 3D UNet was a common choice for UIA detection and segmentation \cite{Timmins2021}, the newer Vision Transformer (ViT) and its variants (e.g., Swin Transformer) that leverage self-attention mechanisms \cite{Vaswani2017AttentionIA,LiuSWin2021} have become increasingly popular in medical imaging applications, providing superior performance. More recently, to better encode contextual information with a lighter model, Yang et al. \cite{Yang2022FocalMN} proposed focal modulation for DL in vision tasks, and demonstrated that it performed better than the Swin Transformer. In focal modulation networks, self-attention is replaced by a stack of depth-wise convolutional layers that focally encode visual contexts and selectively gather them into a modulator using a gated aggregation. Next, the modulator is injected into the query and passed to the next block. Both self-attention and focal modulation techniques involve linearly determining the key, query, and value based on input tokens. Nevertheless, the primary distinction lies in their operational sequence: in self-attention, a query-key interaction occurs before aggregation with the value, whereas in focal modulation, the value is first aggregated with contexts around each key, and subsequently, the result is adaptively modulated with the query. Despite excellent performance in classification and segmentation of natural images, focal modulation's performance in 3D medical images is yet to be verified. Recently, Naderi et al. \cite{naderi2022focal} proposed a 2D UNet-like architecture with focal modulation blocks as the encoder and decoder for abdominal CT segmentation. It offered a higher mean Dice score (not confirmed by statistical tests) than the Swin-Unet \cite{cao2023swin}, another UNet-like model with Swin Transformers as the encoder and decoder. Different from \cite{naderi2022focal}, we followed the approach of the recent 3D Swin-UNETR \cite{hatamizadeh2022swin}, a popular Swin-CNN hybrid model to build a new architecture by replacing the Swin Transformer with 3D focal modulation for the encoder. At the encoder branch, each layer's output, consisting of image tokens and their embeddings, is reshaped to a volumetric patch and passed through a residual block to work as skip connections. Starting from the bottleneck, each feature map is expanded using a transposed convolutional block and then concatenated with the corresponding skip connection before passing through another residual block to the next layer. The detailed architecture is depicted in Fig.~\ref{focalconvunet}. 
        
        
        
    
    \subsection{Post-processing with fully connected CRF}
        \label{sec:methodology:postproc}
        As the proposed deep learning (DL) models (i.e., FocalSegNet) were trained on rough segmentations that overestimate the true aneurysm shapes, we used a fully connected CRF \cite{krahenbuhl2011efficient} to further refine our model's initial prediction ($P(x_i)$). If we define $X_i$ as a random variable representing the assigned label to pixel $i$ (foreground/background), and $I_i$ as a global observation characterizing the pixel information in the input image, the CRF is modeled as the pair of $(I, X)$ which follows a Gibbs distribution of the form $P(X=x|I)=\frac{1}{Z(I)} exp(-E(x|I))$ \cite{lafferty2001conditional}. Here, $Z(I)$ is a normalization factor, and $E(x|I)$ is the energy function of assigning labels $x$ to the image. Through minimizing this energy function, the new labels will have a greater likelihood of being assigned to the image. Inspired by \cite{chen2017deeplab}, we define the energy function as:
        \begin{equation}
            E(x|I)=\sum_i{\psi_u (x_i)} + \sum_{i<j}{\psi_p (x_i, x_j)} 
        \end{equation}
        \begin{equation*}
            \psi_u (x_i)=-log P(x_i)
        \end{equation*}
        \begin{equation*} \resizebox{\textwidth}{!}{$\psi_p (x_i, x_j)=\mu(x_i, x_j)\left [ \omega_1 exp(-\frac{{||p_i - p_j||}^2}{2\sigma_\alpha^2} -\frac{{||I_i - I_j||}^2}{2\sigma_\beta^2}) \\
            + \omega_2 exp(-\frac{{||p_i - p_j||}^2}{2\sigma_\gamma^2}) \right ]$}
        \end{equation*}
        \begin{equation*}
            \mu(x_i, x_j)=1 \; \text{if}\; x_i \neq x_j \;\text{and}\; 0 \;\text{otherwise}
        \end{equation*}
        Here, $\psi_u (x_i)$ is the unary term measuring the cost of assigning label $x_i$ to pixel $i$, and $\psi_p (x_i, x_j)$ is the pairwise term that shows the cost of assigning labels $x_i$ and $x_j$ to pixels $i$ and $j$ simultaneously. The unary term is derived from our FocalSegNet output logits, and the pairwise term is computed by applying a bilateral kernel and a Gaussian kernel on the original MRA patch image. The first kernel forces the pixels with similar positions ($p$) and intensities ($I$) to have similar labels, and the second enforces the smoothness of prediction by only considering the spatial proximity. $\omega_1$ and $\omega_2$, $\sigma_\alpha$, $\sigma_\beta$, and $\sigma_\gamma$ are hyper-parameters that control the weight and scale of these kernels and are chosen empirically. We used connected-component analysis \cite{silversmith_william_cc3d_2021} to identify different clusters when the algorithms predict multiple UIAs, and each connected component was treated as an individual detected aneurysm. To further remove noise, any cluster that had lower than 10 voxels was discarded.

\section{Experiments and Evaluation}
    \label{sec:experiment}
    \subsection{Dataset and pre-processing}
        \label{sec:experiment:dataset}
        Unfortunately, we were not given access to the ADAM challenge dataset \cite{Timmins2021}. Instead, we relied on a publicly accessible TOF-MRA dataset of UIAs described in \cite{Noto2023} to develop and validate our algorithms. It contains 284 subjects (170 females, 127 healthy controls/157 patients with 198 aneurysms). For 246 subjects, coarse spheres that enclose a whole aneurysm were manually labeled, while 38 subjects have voxel-wise segmentation. All scans were pre-processed with skull-stripping and bias field correction \cite{RN13}, and finally resampled to a median resolution of $0.39\times0.39\times0.55$ $mm^{3}$. To allow efficient computation, we performed segmentation based on 3D image patches instead of the entire brain volume. Furthermore, we followed the ``anatomically informed" approach in \cite{Noto2023} to extract image patches of $64\times64\times64$ voxels guided by a probabilistic cerebrovascular atlas with anatomical landmarks on the most probable locations of having an aneurysm, resulting in $\sim$50 patches per subject. In summary, several positive patches with different offsets around every aneurysm were extracted, and for negative patches, a variety of vessel-like, landmark-centered, and random patches were obtained. For more details and the script for image patch extraction, we refer the readers to the original data paper \cite{Noto2023}. It is worth mentioning that we used the same patch-extracting strategy for train and test sets. Finally, for DL model training and inferencing, all image patches were normalized with z-transformation. Note that the cases with coarse labels were divided subject-wise for training (95\%) and validation (5\%), and those with refined segmentation masks were saved for model testing only. 

    \subsection{Implementation details and baseline models}
        \label{sec:experiment:implementation}
        In our proposed FocalSegNet, we utilized a 3D version of the original Focal modulation network \cite{Yang2022FocalMN} as the encoder. Each layer of the encoder consists of two Focal Modulator Blocks, comprising hierarchical depth-wise convolutional layers with kernel sizes of 3, 5, and 7, alongside GELU activation functions. For the baseline models to compare against the FocalSegNet, we also implemented a 3D Residual-UNet that unlike the simple UNet used in the original data paper \cite{Noto2023}, takes advantage of residual connections for better gradient and information flow, and a Swin-UNETR with default parameters \cite{hatamizadeh2022swin} to compare two similar mechanisms, self-attention and focal modulation. The only difference between Swin-UNETR and FocalSegNet is their encoders, which makes our comparison valid. All networks had four layers/hierarchies, equal embedding dimensions, and were trained with a batch size of 12 and an AdamW optimizer (weight decay for L2 regularization=1e-6 and initial learning rate=1e-3). We used a step learning scheduler that reduces the learning rate by 2\% every 100 iterations. Furthermore, we used sampling, data augmentation, and a combination of different loss functions to tackle the class imbalance problem. Early stopping is also used to avoid model overfitting.
                
        \subsubsection{Sampling and data augmentation:}
            \label{sec:experiment:augmentation}
            The ratio of patches without and with UIAs is 9:1, which can cause the model to be biased toward the negative class. To tackle this, each patch in the training set is assigned a probability distribution inversely proportional to its label's frequency of occurrence. Then, each batch is randomly sampled from this distribution with replacement, ensuring the number of positive and negative patches in a batch is almost the same. Although some samples may not be seen in one epoch, there is a high likelihood that the network will see all of them after several epochs. We also use online image augmentation techniques, including random flipping, rotation of up to 15 degrees along x-, y-, and z-directions, and the addition of random Gaussian noise to mitigate overfitting.
            
        \subsubsection{Loss function}
            \label{sec:experiment:loss}
            Since the small sizes of UIAs can cause a large class imbalance, we composed our total loss function with likelihood-based cross-entropy loss (CE), regional-based generalized Dice loss (GD) \cite{sudre2017generalised}, and distance-based boundary loss (Boundary) \cite{Kervadec2018BoundaryLF} as:
            \begin{equation} 
                    Loss = \alpha CE + \beta GD + \gamma Boundary
            \end{equation}
            \begin{equation*}
                CE = -\sum_c{g_c \cdot log(p_c)}
            \end{equation*}
            \begin{equation*}
                GD = 1 - 2 \times \sum_c{\omega_c\frac{p_c \odot g_c}{p_c + g_c}}
            \end{equation*}
            \begin{equation*}
                Boundary = p^1 \cdot \phi(g^1)
            \end{equation*}
                
            \noindent where g and p are ground-truth and prediction logits of a pixel, respectively. The $\omega$ is a weight assigned to each class inversely proportional to its frequency, and $\phi(g^1)$ is the distance map of the foreground class, and its value at each pixel equals the Euclidean distance between that point and the closest background point. $\alpha=\beta=2\gamma=0.40$ were determined empirically. These individual loss functions help improve the detection and segmentation of UIAs for the proposed and baseline models.

    \subsection{Evaluation metrics}
        \label{sec:experiment:eval_metrics}
        For the outcomes of the proposed algorithm and its counterparts, as well as the associated ablation studies, we included five different metrics to evaluate their performance in detecting and segmenting the aneurysm based on the test dataset on a per-image-patch basis. Specifically, for UIA detection quality, we measured the sensitivity and false positive (FP) rate. Note that within the same image patch, it is feasible to contain multiple UIAs. Thus, we separated distinct connected components in the prediction map and ground truth, and evaluated them per aneurysm. Here, a correct aneurysm detection is defined as when the centroid of the predicted segmentation component lies within the boundary of an aneurysm and is considered a false positive otherwise. Next, for all the correctly identified aneurysms, we further evaluated the segmentation accuracy using the Dice coefficient, Intersection over Union (IoU), and 95-$\%$ Hausdorff distance (95-HD) \underline{measured in voxels} to verify the quality of region and boundary overlaps. Because an aneurysm may be detected by one model and missed by the other, we performed an unpaired two-sided t-test to compare the performance between different experimental setups. A $\text{p-value}<0.05$ was used to indicate a significant difference.

\section{Results}
    \label{sec:results}
    \subsection{UIA detection and segmentation}
        \label{sec:results:uia}
\begin{figure}[ht]
\includegraphics[width=\textwidth]{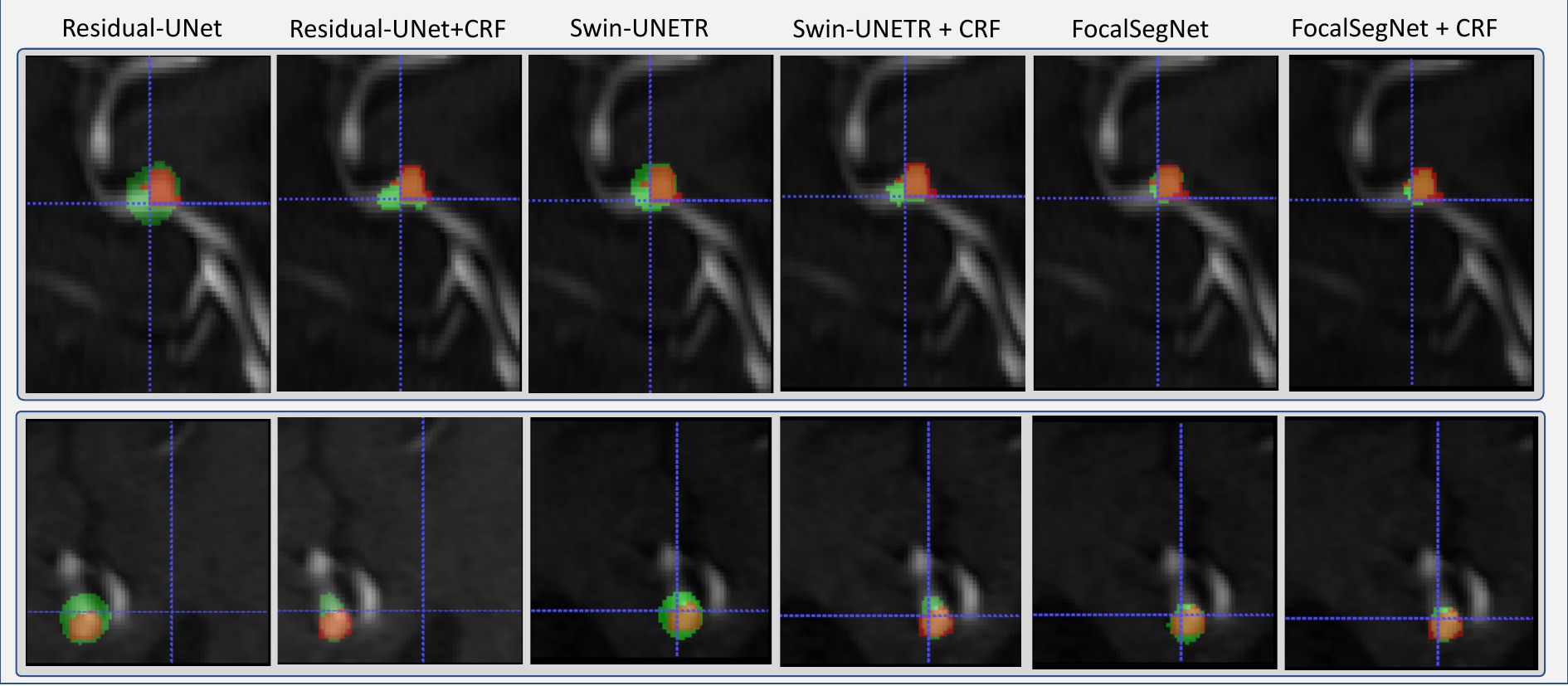}
\caption{Comparison of segmentation results of different techniques for two different patients (one patient per row).Red label=ground truths and green label=automatic segmentation.} \label{qualitative_results}
\end{figure}

        The UIA detection and segmentation performance for all three DL models (i.e., Residual-UNet, Swin-UNETR, and FocalSegNet) with CRF finetuning is listed in Table \ref{tab:results}, with segmentation results for two subjects illustrated in Fig.~\ref{qualitative_results}. In terms of aneurysm detection, for FP rate, our FocalSegNet well outperformed the other two networks ($\text{p}<0.05$), while for sensitivity, all methods perform similarly ($\text{p}>0.05$). This implies that FocalSegNet is better at distinguishing true/false UIAs, which could be hard even for human raters at times. Although FocalSegNet has the best mean segmentation metrics, the difference with Swin-UNETR was not significant ($\text{p}>0.05$). However, both outperform the Residual-UNet ($\text{p}<0.05$). 

\begin{table*}[htbp]
  \centering
  \renewcommand{\arraystretch}{2}
  \caption{UIA detection and segmentation performance (mean±std) of different deep learning models with and without CRF post-processing.}
  \label{tab:results}
  \resizebox{\textwidth}{!}{
    \begin{tabular}{|c|c|c|c|c|c|}
      \hline
         & \textbf{FP rate} & \textbf{Sensitivity} & \textbf{Dice} & \textbf{IoU} & \textbf{95-HD} (voxels)  \\ \hline
			\textbf{Residual-UNet}     		&  0.322 ± 0.537 & 0.793 ± 0.405 & 0.587 ± 0.144 & 0.430 ± 0.140 & 2.870 ± 1.262	\\ \hdashline
			\textbf{Residual-Unet + CRF}	&  0.277 ± 0.500 & 0.778 ± 0.415 & 0.668 ± 0.129 & 0.515 ± 0.137 & 2.315 ± 1.125	\\ \hline 
			\textbf{Swin-UNETR}        		&  0.476 ± 0.596 & 0.866 ± 0.340 & 0.625 ± 0.137 & 0.468 ± 0.137 & 2.495 ± 0.957	\\ \hdashline
			\textbf{Swin-UNETR + CRF}       &  0.403 ± 0.560 & 0.841 ± 0.366 & 0.668 ± 0.144 & 0.518 ± 0.149 & 2.214 ± 1.032	\\ \hline
			\textbf{FocalSegNet}     		&  0.231 ± 0.488 & 0.827 ± 0.379 & 0.638 ± 0.130 & 0.481 ± 0.132 & 2.504 ± 1.140	\\ \hdashline
			\textbf{FocalSegNet + CRF}     	&  0.212 ± 0.464 & 0.801 ± 0.399 & 0.677 ± 0.141 & 0.527 ± 0.144 & 2.148 ± 1.082	\\ \hline
        \end{tabular}%
  }
\end{table*}

    \subsection{Ablation studies}
        \label{sec:results:ablation}
        Besides comparing the performance of the full setup, where fully-connected CRF was used to refine the initial results of the DL models, including FocalSegNet, Swin-UNETR, and Residual-UNet, we also conducted ablation studies to test the impact of CRF post-processing, as well as each component of the full loss function for the proposed FocalSegNet. We evaluated the metrics for UIA detection and segmentation for all the experiments to gain the required insights. 
        
        From Table \ref{tab:results}, we observe that CRF filtering has a positive impact in boosting the accuracy of the initial segmentation of all models ($\text{p}<0.05$) while FocalSegNet offers the best performance without CRF ($\text{p}<0.05$). Table \ref{tab:ablation} demonstrates the impact of design choices in terms of the loss function on FocalSegNet. The network trained solely with the Cross-Entropy loss function performs poorly due to the aneurysm size being negligible compared to background voxels, leading to a bias toward negative predictions. Incorporating Generalized Dice Loss enhances performance, but the false positive (FP) rate remains high. To address this, boundary loss is added, enforcing closer proximity between predictions and ground truths. Ultimately, post-processing with a CRF further enhances segmentation performance.

    \begin{table*}[htbp]
    \centering
    \renewcommand{\arraystretch}{2}
    \caption{Influence of loss functions and CRF post-processing on the proposed FocalSegNet in UIA detection and segmentation results (mean±std)}
    \label{tab:ablation}
    \resizebox{\textwidth}{!}{
        \begin{tabular}{|c|c|c|c|c|c|}
            \hline
                \multicolumn{1}{|l|}{}           & \textbf{FP rate} & \textbf{Sensitivity} & \textbf{Dice} & \textbf{IoU}  & \textbf{95-HD} (voxels) \\ \hline
        			\textbf{Cross-Entropy loss}      & 0.002 ± 0.048    & 0.006 ± 0.075        & 0.018 ± 0.007 & 0.009 ± 0.004 & 4.215 ± 0.121  \\ \hline
        			\textbf{+ Generalized Dice loss} & 0.527 ± 0.729    & 0.864 ± 0.343        & 0.604 ± 0.143 & 0.447 ± 0.144 & 2.854 ± 1.291  \\ \hline
        			\textbf{+ Boundary loss}         & 0.231 ± 0.488    & 0.827 ± 0.379        & 0.638 ± 0.130 & 0.481 ± 0.132 & 2.504 ± 1.140  \\ \hline
        			\textbf{+ CRF}                   & 0.212 ± 0.464    & 0.801 ± 0.399        & 0.678 ± 0.141 & 0.527 ± 0.144 & 2.148 ± 1.082  \\ \hline
        \end{tabular}%
    }
    \end{table*}

\section{Discussion}
    \label{sec:discussion}
    Cerebral aneurysm segmentation is still challenging due to its high similarity in adjacent vessel structures and typically small sizes in contrast to the full brain volume. As previous reports \cite{Noto2023,Timmins2021} primarily focus on binary detection and/or localization of cerebral aneurysms, the relevant reports on the assessment of segmentation algorithms are very limited. Many earlier primary works rely on contrast-enhanced CT angiographies, which have sharper contrast for vasculatures than TOF-MRA, and in-house collected data with refined annotations. Compared with the best segmentation results from the ADAM Challenge (for detected true UIAs) with a 0.64 mean Dice score and 95-HD of 2.62 mm \cite{Timmins2021}, our proposed FocalSegNet, together with CRF post-processing has achieved a Dice score of 0.677$\pm$0.141 and a 95-HD of 2.15 voxel ($\sim$0.95mm), only by leveraging rough voxel-wise annotations of the aneurysms. Regarding UIA detection/localization, our proposed method obtained a sensitivity of 0.801 and an FR rate of 0.212, which are also excellent results. Note that since our UIA detection performance was assessed on a per image-patch basis, the resulting metrics may not be directly comparable to some of those previous reports that were done subject-wise. 
    
    In almost all previous reports on the relevant topic, the UNet and its variants have been widely adopted. Here, we employed the 3D Residual-UNet as a baseline model to properly assess the segmentation quality against the FocalSegNet and Swin-UNETR, whose general architectures were also inspired by the UNet, but with the encoding branch modified with Transformer or focal modulation blocks, respectively. This comparison also allows us to probe the characteristics and performance of these two new DL techniques that model long-range contextual information through different mechanisms. For our application in a 3D weakly supervised segmentation, we confirm the claim of \cite{Yang2022FocalMN} for the benefit of focal modulation, with Swin-UNETR ranking the second. This is likely because aneurysms, especially the small ones, are often attached to the main arteries near the branching points, and thus contextual knowledge of anatomy will be beneficial. This is in addition to the anatomical prior-based approach that we adopted from Di Noto et al. \cite{Noto2023} in training data sampling. 
    
    In terms of GPU memory usage during training, FocalSegNet, Swin-UNETR, and Residual-UNet took 11.2GB, 17.4GB, and 9.7GB respectively, based on the batch size of 12. For all networks, the inference time per patch is around 70ms based on a desktop computer with Intel Corei9 CPU, Nvidia GeForce RTX 3090 GPU, and 24GB RAM. This offers a glance of the efficiency of the proposed network. Compared to \cite{Yang2022FocalMN,naderi2022focal}, we further extended the 2D focal modulation to 3D in a new UNet-like architecture for segmentation tasks for the first time. In future works, we will further examine the proposed FocalSegNet for other supervised and weakly supervised segmentation tasks in other anatomical structures.
    
    For all the DL models, we used a combination of different loss functions, including cross-entropy loss, generalized Dice loss, and boundary loss. While generalized Dice loss and cross-entropy loss have been popular for many medical image segmentation tasks, boundary loss \cite{Kervadec2018BoundaryLF} is helpful in segmenting structures with high-class imbalance, such as the case of aneurysms segmentation. With the intention to reduce the reliance on carefully annotated manual segmentation, many have attempted different strategies in weakly or semi-supervised learning \cite{chan2021comprehensive}, among which CRFs have often been used either in pre-processing pseudo ground truth labels or post-processing results from DL models trained on weak labels. In our case, we designed the system using a fully connected CRF model as a post-processing step, which further improved the accuracy of the initial automatic segmentation based on the spatial and intensity consistency of the aneurysms. Some other works have also reported the application of a subsequent refinement neural network \cite{adiga2020manifold}, but this approach often requires a small number of refined manual segmentations and will lead to a semi-supervised framework instead.   

\section{Conclusion}
    \label{sec:conclusion}
    In conclusion, we have proposed a novel 3D focal modulation UNet called FocalSegNet in combination with CRF for weakly supervised segmentation and detection of brain aneurysms from TOF-MRA. By leveraging coarse segmentation ground truths, the proposed technique was able to achieve excellent performance. By comparing its performance with the popular Residual-UNet and the most recent Swin-UNETR, we demonstrated its superior performance and will extend its application to other domain tasks in the near future.

\section*{CRediT authorship contribution statement}
\textbf{Amirhossein Rasoulian:} Conceptualization, Methodology, Software, Validation, Investigation, Writing - Original Draft, Writing - Review \& Editing, Visualization. \textbf{Arash Harirpoush:} Conceptualization, Software. \textbf{Soorena Salari:} Conceptualization, Writing - Review \& Editing, Formal analysis. \textbf{Yiming Xiao:} Conceptualization, Writing - Review \& Editing, Supervision, Project administration.

\section*{Declaration of competing interest}
The authors declare that they have no known competing financial interests or personal relationships that could have appeared to
influence the work reported in this paper.

\section*{Data availability}
The dataset used in this project is publicly available at \url{https://openneuro.org/datasets/ds003949/versions/1.0.1}.

\section*{Acknowledgments}
This work was supported by a Fonds de recherche du Québec – Nature et technologies (FRQNT) Team Research Project Grant (2022-PR296459).

\section*{Declaration of Generative AI and AI-assisted technologies in the writing process}
During the preparation of this work the authors used ChatGPT in order to improve readability and language style. After using this tool, the authors reviewed and edited the content as needed and take full responsibility for the content of the publication.



 \bibliographystyle{elsarticle-num} 
 \bibliography{elsarticle-template-num}





\end{document}